\documentclass[traditabstract]{aa}
\usepackage{epsfig}

\def\gtsima{$\; \buildrel > \over \sim \;$}
\def\ltsima{$\; \buildrel < \over \sim \;$}
\def\gtrsim{\lower.5ex\hbox{\gtsima}}
\def\lesssim{\lower.5ex\hbox{\ltsima}}

\begin{document}

\title{A disrupted bulgeless satellite galaxy as counterpart of the ultraluminous X-ray source ESO 243-49 HLX-1}

\author{M. Mapelli\inst{1}, F. Annibali\inst{2}, L. Zampieri\inst{1}, R. Soria\inst{3}}
\institute{INAF-Osservatorio Astronomico di Padova, Vicolo dell'Osservatorio 5, I--35122, Padova, Italy\\ \email{michela.mapelli@oapd.inaf.it}
\and
INAF-Osservatorio Astronomico di Bologna, Via Ranzani 1, I--40127 Bologna, Italy
\and
International Centre for Radio Astronomy Research, Curtin University, GPO Box U1987, Perth, WA 6845, Australia
}
\titlerunning{A bulgeless satellite as counterpart of HLX-1}
 
\authorrunning{Mapelli et al.}
 
\abstract{
The point-like X-ray source HLX-1, close to the S0 galaxy ESO 243-49, is one the strongest intermediate-mass black hole candidates, but the nature of its counterpart is still puzzling. By means of N-body/smoothed particle hydrodynamics simulations, we investigate the hypothesis that the HLX-1 counterpart is the nucleus of a bulgeless satellite  galaxy, which undergoes a minor merger with the S0 galaxy.
We derive synthetic surface brightness profiles for the simulated counterpart of HLX-1 in six {\it Hubble Space Telescope} ({\it HST}) filters, ranging from far ultraviolet (FUV) to infrared wavelengths, and we compare them with the observed profiles. 
Our model  matches the  emission associated with the HLX-1 counterpart in all considered filters, including the bluer ones, even without requiring the contribution of an irradiated disc. The simulation can also account for an extended FUV emission, of which there are some hints from the analysis of the {\it F140LP} {\it HST} filter. This matching is impossible to achieve by assuming either a bulgy satellite, or a young star cluster, or an irradiated disc component.
}
\keywords{
galaxies: interactions -- methods: numerical -- galaxies: individual: ESO 243-49 -- X-rays: individual: HLX-1
}

\maketitle

%

\section{Introduction}~\label{sec:intro}
The X-ray source 2XMM~J011028.1$-$460421 (hereafter HLX-1, Farrell et al. 2009) is one of the strongest intermediate-mass black hole (IMBH) candidates. The peak X-ray luminosity ($\sim{}10^{42}$ erg s$^{-1}$), combined with a disk-black body spectrum peaking at $T_{\rm in}\sim{}  0.2$ keV, 
indicates a black hole (BH) mass $\sim{}10^4$ M$_\odot{}$ (Farrell et al. 2009; Davis et al. 2011; Servillat et al. 2011; Godet et al. 2012). In particular, the X-ray light curve shows a periodic `fast rise exponential decay' (FRED) behaviour, with a semi-regular periodicity of $\sim{}370$ days. This has been interpreted as the orbital period of the companion star (Lasota et al. 2011; Soria 2013). 
The recently observed transient radio emission further supports the IMBH hypothesis (Webb et al. 2012). HLX-1 is projected in the sky at $\sim{}0.8$ kpc out of the plane and $\sim{}3.3$ kpc away from the nucleus of the S0/a galaxy ESO~243-49 (luminosity distance $\sim{}96$ Mpc). The galaxy ESO~243-49 is a member of the cluster Abell~2877 (e.g. Malumuth et al. 1992). The  vicinity of HLX-1 with ESO~243-49 is confirmed by the redshift of the observed H$\alpha{}$ emission line of the counterpart (Wiersema et al. 2010), although the velocity offset between this and the bulge of  ESO~243-49 is $\approx{}400$ km s$^{-1}$, close to the escape velocity from the S0 galaxy (Soria, Hau \&{} Pakull 2013). 

 The optical counterpart of HLX-1  was detected in various bands, from near infrared to far ultraviolet (FUV, Wiersema et al. 2010; Soria et al. 2010, 2012, hereafter S10, S12, respectively; Farrell et al. 2012, hereafter F12), but its nature remains puzzling. In particular, it is still unclear which fraction of the optical/ultraviolet (UV) emission comes from the X-ray-irradiated outer accretion disc, and which fraction from a compact stellar population surrounding the BH (F12, S12). If the contribution from the irradiated disc (ID) is negligible, (at least) a fraction of the stellar population surrounding the BH must be young ($\sim{}10$ Myr), because the optical colours are distinctly blue. On the contrary, if the optical counterpart is dominated by the ID, the stellar population must be old, not to overproduce the blue/UV emission. 

Three main scenarios were proposed to explain the nature of the counterpart: (i) the counterpart is a (young or old) star cluster (SC, e.g. S12; F12), (ii) the counterpart is  the  nucleus  of a disrupted satellite galaxy, which is undergoing minor merger with the S0 galaxy (e.g., Bellovary et al. 2010; S10; Webb et al. 2010; Mapelli et al. 2012, hereafter M12; Mapelli et al. 2013, hereafter M13), or (iii) the counterpart is almost entirely due to the ID and the IMBH is naked, apart from the donor star and perhaps a small number of other stars in its sphere of influence (S12; Zampieri et al., in preparation).

The contribution of the ID is predicted to be negligible only in the young SC scenario, while it was found to be crucial in most other proposed scenarios (naked BH, old SC and disrupted bulgy satellite galaxy). M13 showed that even the merger with a bulgy satellite galaxy requires a significant contribution from the ID, as  many pericentre passages are needed to disrupt a stellar bulge, and to make it consistent with the photometry of the HLX-1 counterpart in the infrared. On the other hand, after many pericentre passages the satellite has lost all its gas, and there is no recent SF to contribute to the blue and UV filters.

In this paper, we focus on the minor merger scenario. While in M12 and M13 we assumed that the satellite galaxy has a bulge, in the current paper we simulate a bulgeless gas-rich satellite galaxy. This difference has important consequences, as pure disc satellite galaxies are disrupted faster than discs with bulges (Gnedin, Hernquist \&{} Ostriker 1999). The paper is organized as follows. In Section~2, we summarize the main scenarios proposed to explain the nature of HLX-1. In Section~3, the simulation methods are described. In~Section~4, we discuss the main results. Our conclusions are presented in Section~5.

\section{On the puzzling nature of the HLX-1 counterpart}
 Henceforth, we shall assume that the IMBH in HLX-1 is surrounded by a substantial stellar population (investigating the case of a naked BH is beyond the aim of this paper, and will be considered in a follow-up study). We will compare the two possible scenarios of a massive SC or a disrupted satellite galaxy, and discuss which one is more consistent with the observed optical photometry and at the same time more plausible from an evolutionary point of view.


Either a young ($\approx{}10$ Myr) or an old ($\approx{}10-12$ Gyr) SC are consistent with the available photometry (F12; S12; M13), depending on the contribution of the ID. In the young SC scenario, the presence of an IMBH may be explained with the runaway collapse of massive stars  (e.g. Portegies Zwart \&{} MacMillan 2002), but there are two serious drawbacks: (i) it is difficult to explain the presence of a young SC far from the disc and in an early type galaxy (young SCs are a disc population and are more frequent in late-type galaxies, e.g. Portegies Zwart, MacMillan \&{} Gieles 2011); (ii) if the optical variability of the counterpart (claimed by S12) is confirmed, then a large fraction of the infrared to UV radiation comes from the ID, and the young SC cannot be more massive than $\approx{}10^4$ M$_\odot{}$. It is very difficult to explain the presence of a  $\approx{}10^4$ M$_\odot{}$ IMBH in a $\approx{}10^4$ M$_\odot{}$ SC.  Even if the optical variability is negligible, the mass of a 10-Myr young SC is unlikely to be $>10^5$ M$_\odot{}$ for a realistic value of the extinction ($A_{\rm V}\approx{}0.18$)\footnote{F12 estimate a young SC mass $4\times{}10^6$ M$_\odot{}$ for $A_{\rm V}\approx{}1.3$, but this extinction is  rather extreme when compared with the hydrogen column density ($\approx{}$ a few $10^{20}$ atoms cm$^{-2}$) derived from disc-black-body fits to the X-ray spectra, and with the line-of-sight extinction $A_{\rm V} = 0.053$.}.

In the old SC scenario, an IMBH might have formed through repeated mergers of stellar-mass BHs (e.g., Miller \&{} Hamilton 2002). The position of the HLX-1 counterpart is consistent with the distribution of globular clusters (which are a halo population). On the other hand, in this case a high (although not unrealistically high) level of disc emission is needed to explain the luminosity in the blue and UV bands. 

Two further problems of the (either young or old) SC scenario are that (i) it is difficult to explain a $\sim{}400$ km s$^{-1}$ relative velocity between the counterpart and the bulge of ESO~243-49, and (ii) we can hardly explain the formation of a $\approx{}10^4$ M$_\odot{}$ IMBH even in a $\sim{}10^6$ M$_\odot{}$ young or old SC (i.e. the maximum mass allowed for the counterpart by observations, F12, S12), as this mass ratio requires a very efficient formation pathway (e.g. the discussion in  Portegies Zwart \&{} MacMillan 2002).

The minor merger scenario removes the latter difficulty, as the IMBH would belong to the low-mass tail of the distribution of super massive BHs (SMBHs), located at the centre of galaxies. There is increasing evidence of (both bulgy and bulgeless) galaxies hosting at their centre SMBHs with mass $\lesssim{}10^5$ M$_\odot{}$ (e.g. Filippenko \&{} Sargent 1989; Filippenko \&{} Ho 2003; Barth et al. 2004; Greene \&{} Ho 2004, 2007a, 2007b; Satyapal et al. 2007, 2008, 2009; Dewangen et al. 2008; Shields et al. 2008; Barth et al. 2009; Desroches \&{} Ho 2009; Gliozzi et al. 2009; Jiang et al. 2011a, 2011b; Secrest et al. 2012; Bianchi et al. 2013). 

Furthermore, both the projected position  and the high relative velocity  of the HLX-1 counterpart are consistent with an ongoing merger. In addition, a recent minor merger is consistent with the presence of prominent dust lanes around the nucleus of the S0 galaxy  (e.g. Finkelman et al. 2010; Shabala et al. 2012) and the evidence of UV emission centred on its bulge (S10), indicating ongoing star formation (SF, e.g. Kaviraj et al. 2009). On the other hand, there is no smocking gun that ESO~243-49 underwent a merger (see the discussion in M13). 




In M12 and M13, we simulated the merger between a bulgy gas-rich satellite galaxy and an S0 galaxy. We found that the simulations are in fair agreement with the integrated magnitude of the HLX-1 counterpart in the different bands only if (i) the merger is in a very late state ($>2.5$ Gyr since the first pericentre passage), (ii) we assume that the ID emission dominates over the stellar component in all the filters but the near infrared one. These two requirements rise from the fact that many pericentre passages are needed to disrupt an old stellar bulge, and to make it consistent with the photometry of the HLX-1 counterpart in the infrared. On the other hand, after many pericentre passages the satellite has lost all its gas, and there is no recent SF to contribute to the blue and UV filters. 

The main problem of the bulgy satellite scenario is that  it cannot account for the entire emission from the counterpart: an ID component is needed to explain the blue, near UV (NUV) and FUV emission. Furthermore, the merger with a bulgy satellite cannot explain the extended FUV emission, which was recently claimed in M13.

M13 suggested that the FUV emission associated with the HLX-1 counterpart is more extended than implied by the point spread function (PSF), in the {\it Hubble Space Telescope} ({\it HST}) data. The current data do not allow to exclude that the extended emission is connected with the background galaxy at $z=0.03$ (Wiersema et al. 2010), or even with ESO~243-49. In fact, the FUV emission of the counterpart of HLX-1 is consistent with that of a point-like source superimposed to an extended FUV `halo' (see figure 5 of M13). If the extended FUV emission is physically connected with the HLX-1 counterpart, the SC scenario and even the merger with a bulgy satellite galaxy  are ruled out.

In the current paper, we investigate the scenario of a bulgeless satellite, by means of N-body/smoothed particle hydrodynamics (SPH) simulations. Pure disc satellite galaxies are disrupted faster than disc galaxies with bulges, due to their lower central density, which results in a smaller tidal radius (Gnedin et al. 1999; Feldmann, Mayer \&{} Carollo 2008). Thus, we want to check whether the merger with a bulgeless satellite galaxy allows to explain both the integrated magnitude in the bluer filters and the extended FUV emission.

\section{$N-$body simulations}~\label{sec:nbody}

\begin{table}
\begin{center}
\caption{Initial conditions of the $N-$body simulations.}
 \leavevmode
\begin{tabular}{lll}
\hline
Model galaxy properties$^{\rm a}$ & Primary & Secondary \\
\hline
$M_{\rm DM}$ ($10^{11}$ M$_\odot{}$)          & 7.0  & 0.3\\
$M_\ast{}$  ($10^{10}$ M$_\odot{}$)      & 7.0   & 0.15\\
$f_{\rm b/d}$                               &  0.25 & 0 \\
$M_{\rm gas}$  ($10^{8}$ M$_\odot{}$)    & 0 &  1.38 \\
NFW scale radius (kpc) & 6.0 & 3.0\\
Disc scale length (kpc) & 3.7 & 3.0 \\
Disc scale height (kpc) & 0.37 & 0.30 \\
Bulge scale length (kpc) & 0.6 & -- \\
\noalign{\vspace{0.1cm}}
\hline
\end{tabular}
\begin{flushleft}
\footnotesize{$^{\rm a}$ $M_{\rm DM}$ and $M_\ast{}$ are the total DM mass and the total stellar mass of the galaxy, respectively. $f_{\rm b/d}$  is the bulge-to-disc mass ratio. $M_{\rm gas}$ is the total gas mass.} 
\end{flushleft}
\end{center}
\end{table}

\begin{figure}
\center{{
\epsfig{figure=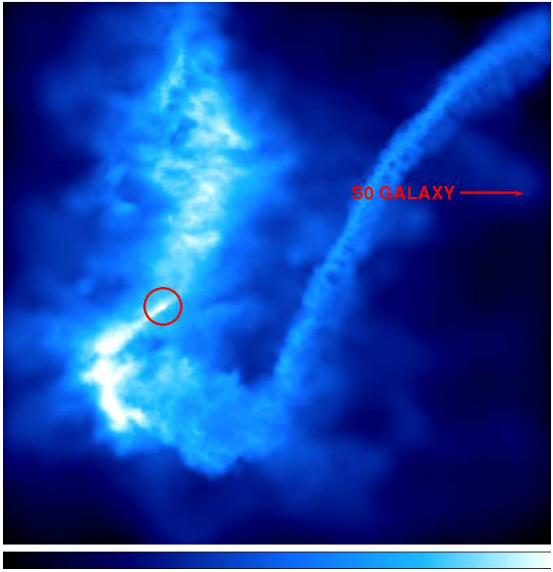,width=7.2cm} 
}}
\caption{\label{fig:fig1}
Colour-coded map: projected mass density of gas in the simulated satellite galaxy at $t=100$ Myr after the first pericentre passage.  The satellite is seen face-on. The scale of the color-coded map is logarithmic, ranging from  $2.2\times{}10^{-9}$ M$_\odot{}$ pc$^{-3}$ (black) to $2.2\times{}10^{-2}$ M$_\odot{}$ pc$^{-3}$ (white).  The frame size is $60 \times 60$ kpc. 
The red circle marks the position of the nucleus of the satellite galaxy. The direction of the S0 bulge is indicated by the red arrow.
}
\end{figure}

\begin{figure}
\center{{
\epsfig{figure=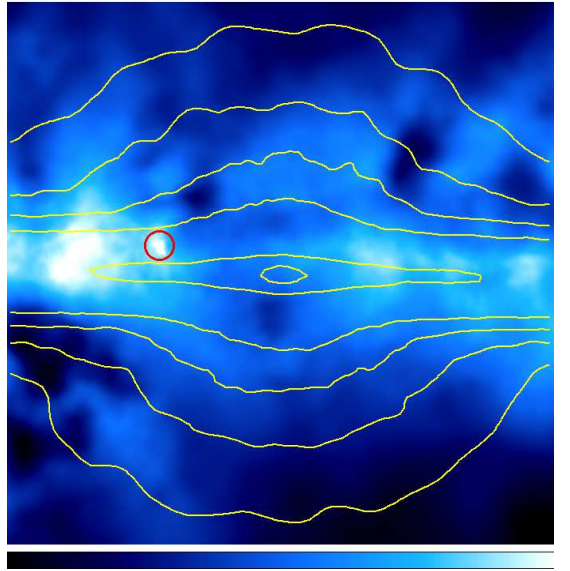,width=7.2cm} 
}}
\caption{\label{fig:fig2}
Colour-coded map: projected mass density of gas in the simulated satellite galaxy at $t=100$ Myr after the first pericentre passage.  Both the satellite and the S0 galaxy are seen edge-on. The scale of the color-coded map is logarithmic, ranging from  $2.2\times{}10^{-9}$ M$_\odot{}$ pc$^{-3}$ (black) to $7.0\times{}10^{-2}$ M$_\odot{}$ pc$^{-2}$ (white).   The frame size is $16 \times 16$ kpc. The red circle marks the position of the nucleus of the satellite galaxy. 
The yellow contours show the projected density of stars in the S0 galaxy. 
}
\end{figure}

As in M12 and M13, the initial conditions for both the primary galaxy and the secondary galaxy in the $N-$body model are generated by using an upgraded version of the code described in Widrow, Pym \&{} Dubinski (2008; see also Kuijken \&{} Dubinski 1995 and Widrow \&{} Dubinski 2005). 
The code generates self-consistent disc-bulge-halo galaxy models, which are very close to equilibrium. In particular, the dark matter (DM) halo is modelled with a Navarro, Frenk \&{} White (1996, NFW) profile. We use an exponential disc model (Hernquist 1993), while the bulge is spherical and comes from a generalization of the S\'ersic law (Widrow et al. 2008). The primary galaxy has a stellar spherical bulge and a stellar exponential disc, and has no gas. The satellite has no bulge, while it has a stellar exponential disc and a gaseous exponential disc. The total mass of the secondary is $\lesssim{}1/20$ of the mass of the primary, classifying the outcome of the interaction as a minor merger. The masses  and the scale lengths of the various components of the simulated galaxies are listed in Table~1. 

The main orbital properties of the interaction are impact parameter $b=10.2$ kpc, initial relative velocity between the centres-of-mass (CMs) of the two galaxies $v_{\rm rel}=50$ km s$^{-1}$, orientation angles $\theta{}=-\pi{}/2$, $\phi{}=0$, $\psi{}=2.94$ rad (for a definition of the angles, see  Hut \&{} Bahcall 1983 and M12), specific orbital energy $E_{\rm s}=-2.03\times{}10^4$ km$^2$ s$^{-2}$, specific orbital angular momentum $L_{\rm s}=0.5\times{}10^3$ km s$^{-1}$ kpc, eccentricity $e=0.999$. The orbit is prograde, i.e. the the orbital angular momentum of the satellite is aligned with the spin of the primary galaxy. 
 The adopted orbits are nearly parabolic, in agreement with predictions from cosmological simulations (Tormen 1997; Wang et al. 2005; Zentner et al. 2005; Khochfar \&{} Burkert 2006; Wetzel 2011). Furthermore, this assumption is consistent with the measured velocity offset between the counterpart of HLX-1 and the bulge of ESO~243-49, which is close to the escape velocity from the S0 galaxy.

The particle mass in the primary galaxy is $10^5$ M$_\odot{}$ and $10^4$ M$_\odot{}$ for DM and stars, respectively.  The particle mass in the secondary galaxy is  $10^4$ M$_\odot{}$ for DM and $7\times{}10^2$ M$_\odot{}$ for both stars and gas. The total number of particles in the simulation is 18.7 Million. The softening length is $\epsilon{}=10$ pc. We also run a test simulation with softening length $\epsilon{}=30$ pc, to make sure that spurious scatterings between particles of different mass are negligible, and we found no spurious effects (see e.g. Bate \&{} Burkert 1997). 

As in M12 and M13, we simulate the evolution of the models with the $N-$body/SPH tree code {\sc gasoline} (Wadsley, Stadel \&{} Quinn 2004). Radiative cooling, SF and supernova blastwave feedback are enabled, as described in Stinson et al. (2009). In particular, SF occurs when cold ($< 3\times{}10^4$ K), virialized gas reaches a threshold density $n_{\rm SF} = 5$ atoms cm$^{-3}$, and is part of a converging flow. SF proceeds at a rate
\begin{equation}
\frac{{\rm d}\rho{}_\ast{}}{{\rm d}t}=\epsilon{}_{\rm SF}\,{}\frac{\rho{}_{\rm gas}}{t_{\rm dyn}}\propto{}\rho_{\rm gas}^{1.5},
\end{equation}
 (i.e. locally enforcing a Schmidt law), where $\rho{}_\ast{}$ and $\rho{}_{\rm gas}$ are the stellar and gas densities, respectively, and $t_{\rm dyn}$ is the local dynamical time. We choose $\epsilon{}_{\rm SF}= 0.1$, in agreement with previous work (e.g. Mapelli \&{} Mayer 2012). 

\section{Results}~\label{sec:hlx1}
\subsection{Gas morphology}
Fig.~\ref{fig:fig1} shows the projected distribution of the gas component of the simulated satellite galaxy at $t=100$ Myr after the first pericentre passage. The satellite galaxy is shown face-on. The two tidal lobes are well evident, as well as the tidal stream connecting the satellite with the bulge of the primary. A circle marks the position of the nucleus of the satellite galaxy. Fig.~\ref{fig:fig2} shows again the projected distribution of the gas component of the simulated satellite galaxy at $t=100$ Myr, but now the satellite is seen edge-on, and rotated to match the observed position of the HLX-1 counterpart with respect to the S0 galaxy.

Figs~\ref{fig:fig1} and \ref{fig:fig2} indicate that the distribution of the atomic hydrogen close to ESO~243-49 should show some interesting features, if the counterpart of HLX-1 is the nucleus of a recently disrupted satellite. 
We notice that the projected distribution of the gas in Fig.~\ref{fig:fig2} qualitatively recalls the distribution of FUV emission in fig.~4 of M13, although the two figures cannot be directly compared. The main difference is that while the FUV emission has a peak in the bulge of ESO~243-49, there is no concentration of gas at the centre of the S0 galaxy in our simulation. This happens because the gas stripped from the satellite galaxy did not have enough time to sink to the centre of the primary galaxy: M12 and M13 showed that it takes $\gtrsim{}1$ Gyr (since the first pericentre passage) for the stripped gas to reach the nucleus of the S0 galaxy.

 The absence of gas at the centre of the simulated S0 galaxy might indicate that the SF observed at the centre of ESO~243-49 is due to residual gas bound to the S0 galaxy, rather than to the gas stripped from the satellite (e.g. Temi, Brighenti \&{} Mathews 2009; M13). 
\begin{figure*}
\center{{
\epsfig{figure=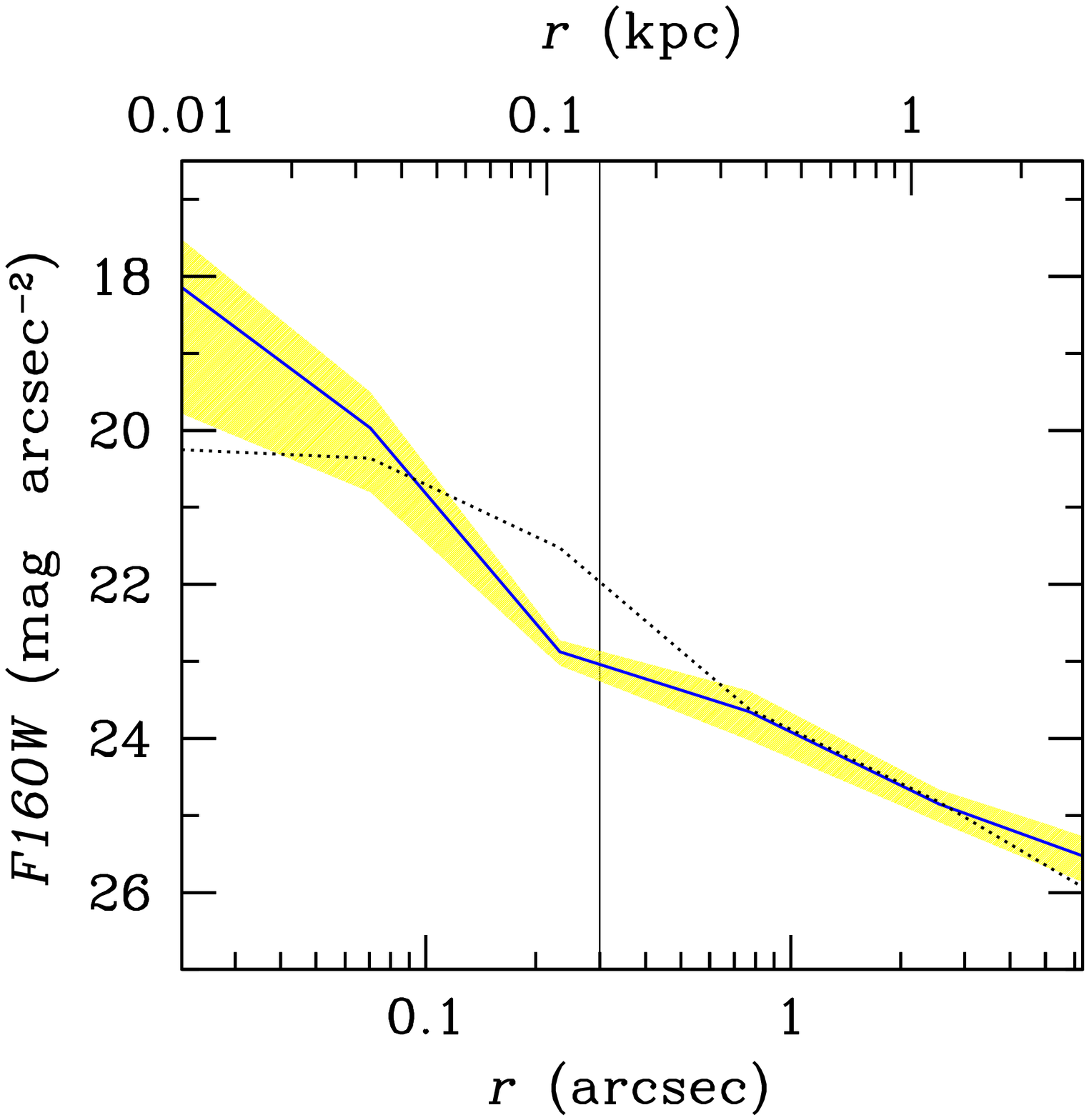,width=7.5cm} 
\epsfig{figure=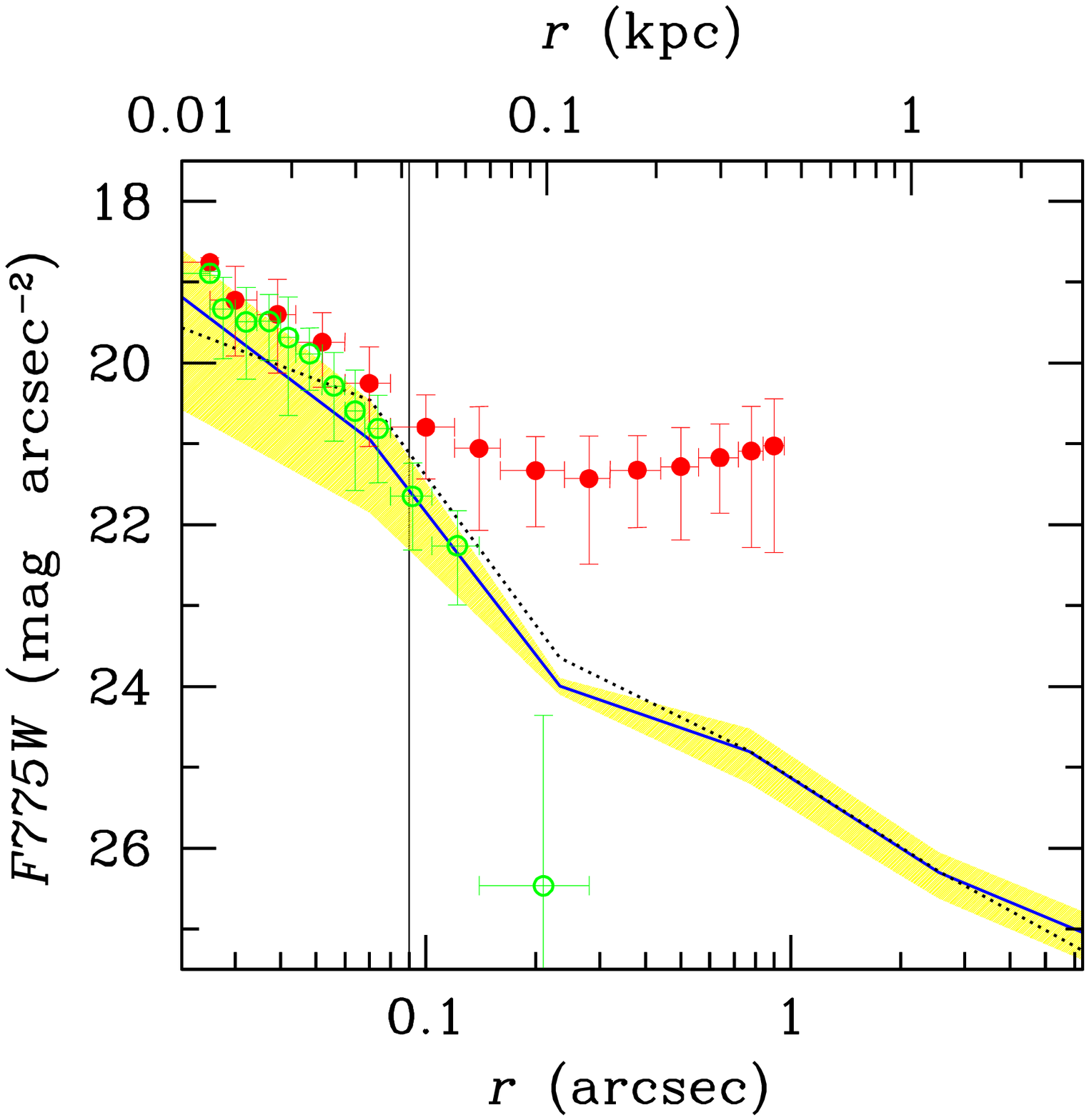,width=7.5cm} 
\epsfig{figure=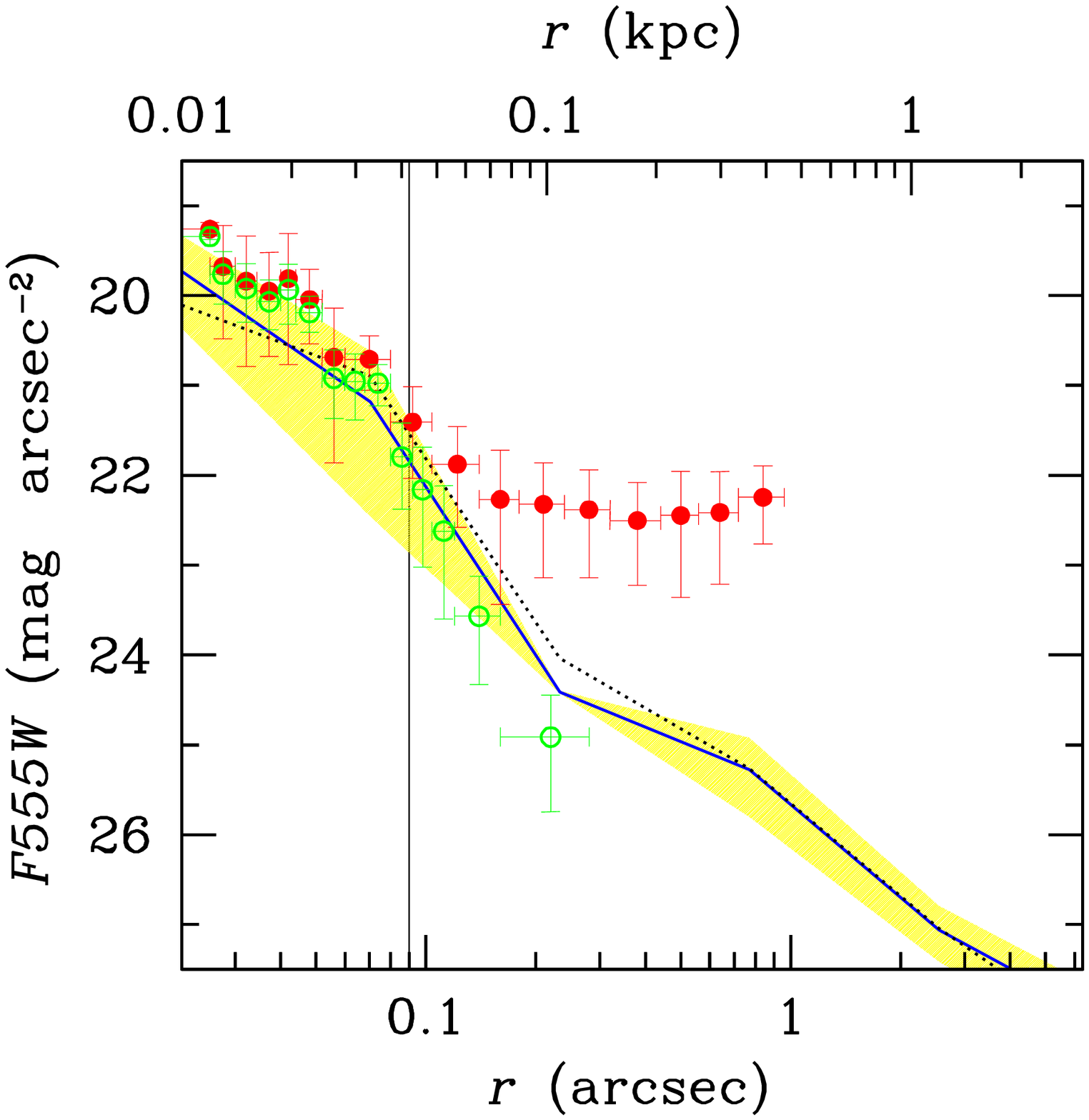,width=7.5cm} 
\epsfig{figure=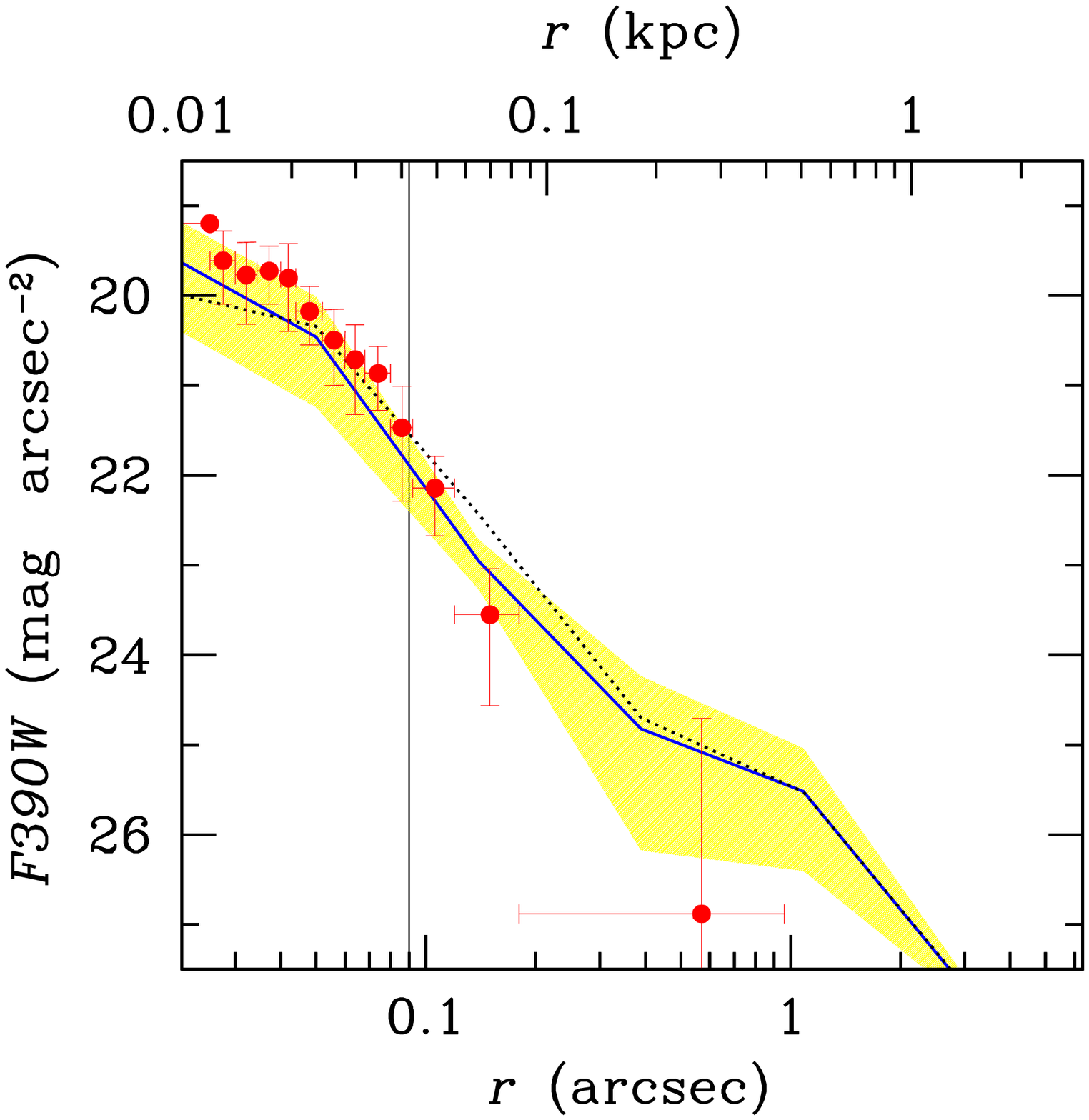,width=7.5cm} 
\epsfig{figure=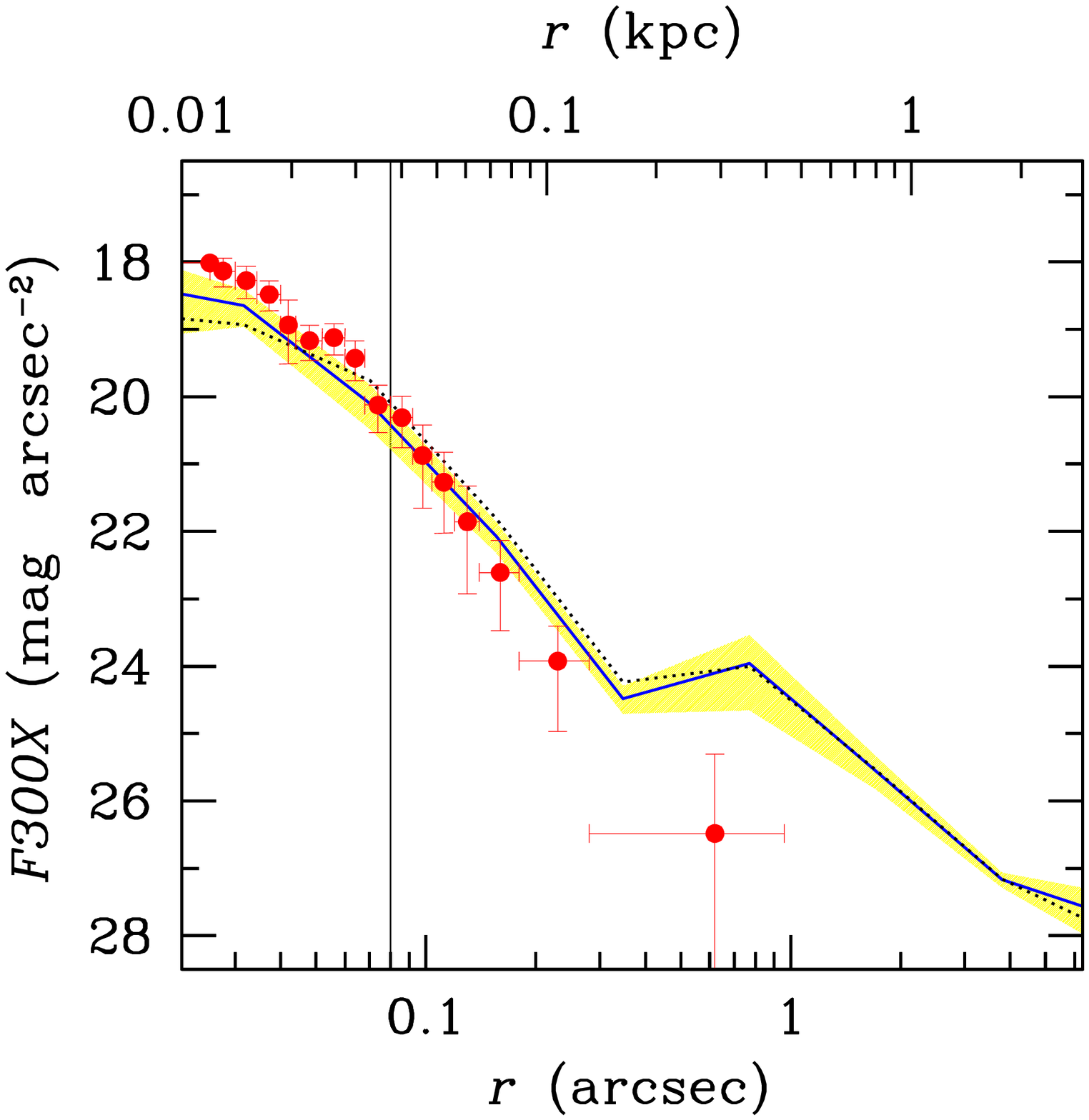,width=7.5cm} 
\epsfig{figure=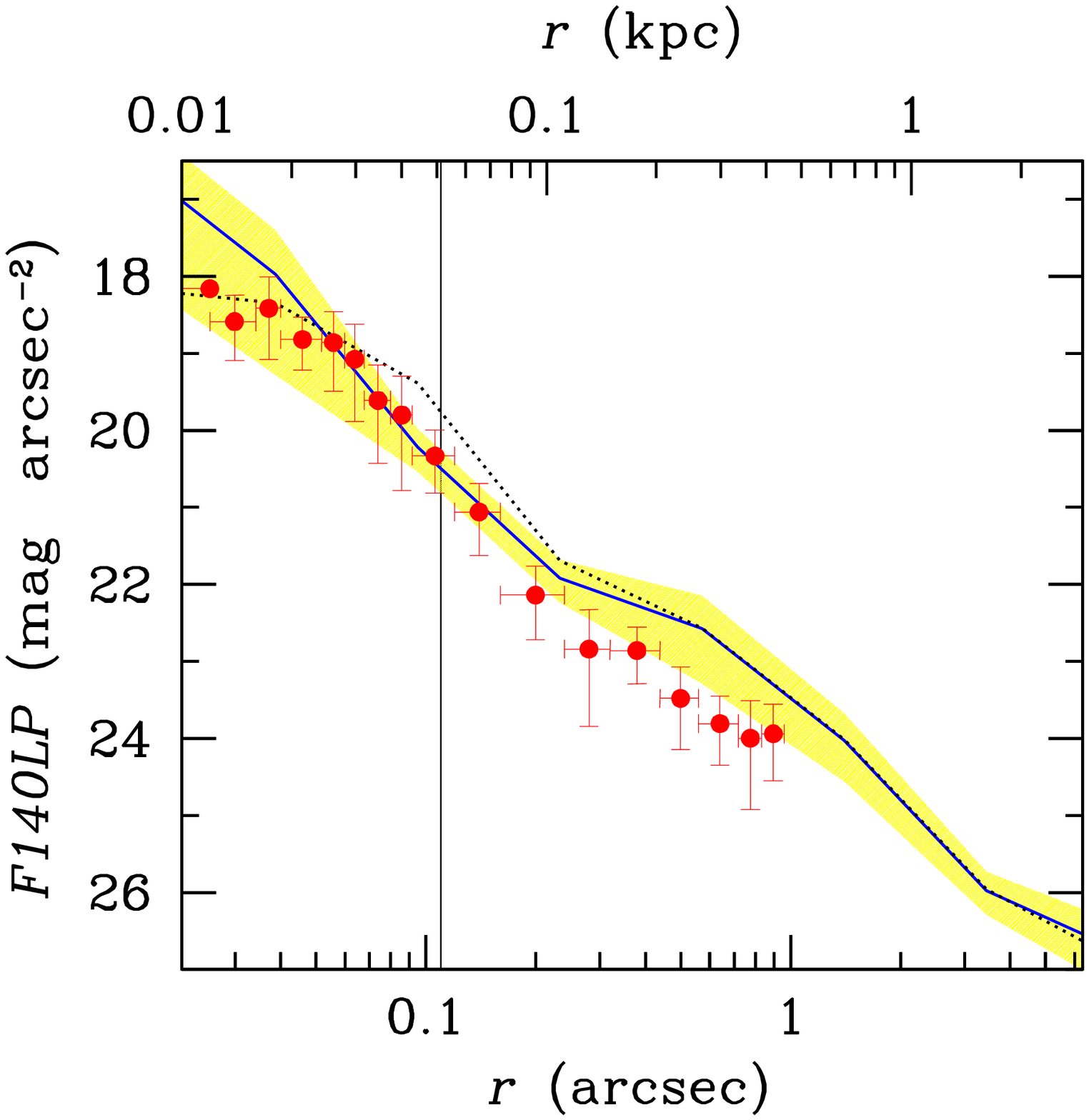,width=7.5cm} 
}}
\caption{\label{fig:fig3}
Surface brightness profiles of the HLX-1 counterpart in the simulation (lines) and in the {\it HST} observations (circles). From left to right and from top to bottom: filter {\it F160W}, {\it F775W}, {\it F555W}, {\it F390W}, {\it F300X} and {\it F140LP}. In all panels, solid blue line: simulated profile of the satellite galaxy at $t=100$ Myr after the first pericentre passage. We assume A$_V=0.18$ for the simulation. The dotted black line is the simulated profile convolved with the observed PSF. The yellow shaded area shows the uncertainty in the simulation due to stochastic fluctuations and to spatial resolution (see the text for details).  The filled red circles were obtained in M13 by subtracting the background according to the first approach (as described in M13). In the {\it F775W} and {\it F555W} filters, the open green circles were obtained by subtracting the background according to the second approach, as described in M13. We do not show the observed profile in {\it F160W} as the background of the S0 galaxy dominates over the flux of the HLX-1 counterpart in this filter.
 The error bars of the data points are at 1 $\sigma{}$. Vertical solid line: PSF full-width at half maximum, as derived in M13.
}
\end{figure*}

\subsection{The stellar component and the surface brightness profiles}

Part of the gas was converted into young stars by applying the SF recipes described in Section~\ref{sec:nbody}. We find that the simulated mass of young stars in the inner 0.4 arcsec of the satellite is $M_{\rm y}(<0.4'')=2.5\times{}10^4$ M$_\odot{}$ at $t=100$ Myr, corresponding to $\sim{}6$ per cent of the total stellar mass within 0.4 arcsec ($=3.9\times{}10^5$ M$_\odot{}$).






\begin{table}
\begin{center}
\caption{Simulated magnitudes within 0.4 arcsec, compared with the observed magnitude within 0.4 arcsec, $m(<0.4'')$.}
 \leavevmode
\begin{tabular}{llll}
\hline
 Filter & Model     & Model   & $m(<0.4'')$\\ 
        & (A$_V=0$) & (A$_V=0.18$) & \\ 
\hline 
 {\it F140LP} & $21.9$ & $22.2$ & $22.21 \pm 0.03$ \\
 {\it F300X}  & $23.0$ & $23.3$ & $22.80 \pm 0.05$\\
 {\it F390W}  & $24.1$ & $24.3$ & $24.04 \pm 0.05$\\
 {\it F555W}  & $24.4$ & $24.6$ & $24.11 \pm 0.05$\\
 {\it F775W}  & $24.0$ & $24.2$ & $23.64 \pm 0.15$\\
 {\it F160W}  & $23.2$ & $23.3$ & $23.49 \pm 0.26$ \\
\noalign{\vspace{0.1cm}}
\hline
\end{tabular}
\begin{flushleft}
\footnotesize{All the reported magnitudes are in Vegamag. In columns 2 and 3 we show the magnitude of the fiducial simulation, assuming A$_V=0$ and A$_V=0.18$ (M13, S12), respectively. The uncertainty in the simulated magnitude is $\approx{}0.5$ mag. In column 4 we report the observed magnitude within 0.4 arcsec, from table~1 of M13, to facilitate the comparison.}
\end{flushleft}
\end{center}
\end{table}

We then derive synthetic fluxes (in the six {\it HST} filters) by using the single stellar population (SSP) models based on the tracks of Marigo et al. (2008), with the Girardi et al. (2010) case A correction for low-mass, low-metallicity asymptotic giant branch stars\footnote{\tt http://stev.oapd.inaf.it/cgi-bin/cmd\_2.3}.  For simplicity, we assume that all the stars already present in the initial conditions of the simulation have an age $t_{\rm old}=12$ Gyr, and that all the young stars that formed as a consequence of the merger have an age $t_{\rm y}=10$ Myr. 

The results are shown in Fig.~\ref{fig:fig3} and in Table 2.
In Fig.~\ref{fig:fig3}, we compare the synthetic surface brightness profiles with the observed surface brightness profiles of the HLX-1 counterpart (derived in M13). We remind that one of the major issues about the optical counterpart of HLX-1 (as highlighted in F12, S12, M12 and M13) is the subtraction of the light coming from the ESO~243-49. The contamination from ESO~243-49 is particularly strong in the infrared ({\it F160W}), where it is impossible to recover a surface brightness profile for the HLX-1 counterpart, and it is still dominant in the {\it F775W} and in the {\it F555W} filters, while it is less critical in the bluer filters. In M13, we used and compared two approaches to remove the background. In the first approach, we computed the background in an annulus\footnote{To account for the uncertainties due to the highly variable S0 background, we repeated the computation varying the position of the background annulus between 1.0 and 1.4 arcsec in steps of 0.2 arcsec, and then averaging among the results (see M13).} of  width 0.08 arcsec at $r>1$ arcsec. In the second approach, we created  a Gaussian smoothed image of the S0 galaxy and then subtracted it to the original frame. The first approach is more conservative and does not smear out possible faint irregular features. On the other hand, the first approach might fail in removing the contribution of the S0 galaxy at large radii ($r>0.2$ arcsec). The first approach works fine for the bluer filters ({\it F140LP}, {\it F300X} and {\it F390W}), but it overestimates the integrated light of the HLX-1 counterpart in the redder ones ({\it F555W} and {\it F775W}) with respect to previous work (F12 and references therein). Thus, even if the level of contamination from ESO~243-49 is still a matter of debate, in M13 and in this paper we take the results of the second approach (green open circles) as reference values for the {\it F555W} and {\it F775W} filters.

The solid line  in Fig.~\ref{fig:fig3} shows our fiducial simulation.  The dotted line is the fiducial run convolved with the observed PSF\footnote{As in M13, we approximate the PSF as a two-dimensional Gaussian with the same full-width at half maximum as the observed PSF.}. We warn that the convolution may lead to underestimate the flux of the simulation in the inner bins, where the simulation is dominated by softening effects. Thus, we report both the non-convolved (solid line) and the convolved (dotted line) profile to give an idea of the upper and lower case for the inner bins. We notice that the effect of the PSF convolution is dramatic only in the case of the {\it F160W} filter, where the PSF full-width at half maximum is $\gtrsim{}3$ times the value of the other filters.

 The yellow shaded area in Fig.~\ref{fig:fig3} shows the uncertainty of the simulation, due to the combined effects of stochastic fluctuations and spatial resolution. 
In particular, it accounts for the differences between the fiducial simulation and a test run, which differ only for the softening length  ($\epsilon=10$ and 30 pc in the fiducial simulation and in the test run, respectively, see Section 3). As we did not find any significant discrepancy in the energy distribution of particles between the fiducial and the test run, we can consider the differences between the two runs at $r>30$ pc as purely stochastic fluctuations, due to the low number of particles in the nucleus of the satellite. Thus, the uncertainty of the simulation (represented by the shaded area) is larger (i) at $r<30$ pc because of the different softening length, and (ii) at very large radii ($\gtrsim{}0.2$ kpc) because of the low number of particles, which increases the effect of stochastic fluctuations.

 The simulation matches the observed profile of the HLX-1 counterpart in all the considered filters, even the bluer ones ({\it F390W}, {\it F300X} and {\it F140LP}), without requiring any contribution from an ID component. The simulation also matches the extended profile of the HLX-1 counterpart in the FUV filter. As shown in M13, this result is impossible to achieve by assuming either a bulgy satellite or an ID component. Thus, the minor merger with a bulgeless gas-rich satellite is the only way to explain the extended FUV emission, if this is associated with the HLX-1 counterpart. 

In Table~2, we report the integrated magnitude within 0.4 arcsec estimated from the fiducial simulation, and we compare it with the observational values derived in M13. There is a fair agreement between the simulation and the data in the FUV ({\it F140LP}) and in the infrared ({\it F160W}) filter, while in the other filters there is a larger discrepancy. 

We remark that the synthetic surface brightness profiles shown in Fig.~\ref{fig:fig3} and the integrated magnitudes reported in Table~2  are not an attempt to fit the observational data, as we ran just one simulation and we have no free parameters to play with.  Given the large uncertainty connected with stochastic fluctuations, we do not think that it is meaningful to run a wider grid of simulations, untill more constraints about the orbit and the gas content of the HLX-1 counterpart become available.

Furthermore, we do not add any ID component, but it is likely that some contribution from the accretion disc is there. Adding an ID component to the stellar population component would increase the range of models that match the data, but would imply additional free parameters.

\section{Conclusions}~\label{sec:conclude}
In this paper, we investigate the hypothesis that the HLX-1 counterpart is the nucleus of a bulgeless satellite  galaxy undergoing a minor merger with the S0 galaxy. A bulgeless satellite is disrupted faster (due to its lower central density, Gnedin et al. 1999) than a bulgy satellite. For this reason, the simulation presented in this paper is able to reproduce the photometry of the counterpart of HLX-1 in all the {\it HST} filters, including the bluer ones ({\it F390W}, {\it F300X} and {\it F140LP}). Instead, a simulated bulgy satellite can account only for the emission in the redder filters ({\it F160W} and {\it F775W}, see M13), as the old stars dominate the stellar population. For a bulgy satellite to be consistent with the counterpart of HLX-1, we need to assume that most of the optical and UV emission comes from an ID (M13), while a bulgeless satellite scenario implies very mild or no contribution from the disc.


M13 recently suggested that the FUV emission associated with the HLX-1 counterpart may be extended. This claim and the nature of the extended emission (i.e. whether it is physically connected with the HLX-1 counterpart or with the ESO 243-49 or with a background galaxy) are still debated. Our simulations show that a bulgeless satellite galaxy can match the extended FUV emission surrounding the HLX-1 counterpart. This matching is impossible to achieve by assuming either a bulgy satellite (M13), a young SC, or an ID component. 
If the satellite has a bulge, it takes a long time to be disrupted ($\gtrsim{}2.5$ Gyr since the first pericentre passage), and  the SF in the satellite is quenched by gas stripping, before the flux of the old stellar component becomes consistent with the observed one in the infrared. 

 We chose a realistic set of orbital parameters for the simulation presented in this paper, and we showed that the results match the photometry of the HLX-1 counterpart. We expect that other choices of the orbital parameters (eccentricity, angular momentum, specific energy) and of the properties of the satellite (especially stellar and gas mass) may produce results in agreement with the observations. While it is prohibitive to cover the entire parameter space with sufficiently high-resolution simulations, a number of different orbital parameters and satellite properties deserve to be investigated in a forthcoming study. This will be particularly helpful if more accurate photometric and spectral constraints on the HLX-1 counterpart become available.

We  add no ID component in our model, but it is likely that the disc emission contributes to the observed flux. It is essential to establish whether the HLX-1 counterpart is variable, as this is the only clue to disentangle an ID from the stellar population component. Finally, our simulations show that tidal tails should be evident in the gas component (Figs.~\ref{fig:fig1} and \ref{fig:fig2}), if the HLX-1 counterpart is the nucleus of a bulgeless satellite galaxy. Thus, observations of the atomic hydrogen and of its kinematics can contribute to shed light on the nature of the HLX-1 counterpart.

\section*{Acknowledgments}
We thank the anonymous referee for their comments that improved the manuscript. We also thank the developers of {\sc gasoline} (especially J. Wadsley, T. Quinn and J. Stadel), L.~Widrow for providing us the code to generate the initial conditions, L.~Girardi and A.~Bressan for providing the SSP models, and E. Ripamonti  for useful discussions. To analyze simulation outputs, we made use of the software TIPSY\footnote{\tt http://www-hpcc.astro.washington.edu/tools/tipsy/\\tipsy.html}.
We acknowledge the CINECA Award N. HP10CLI3BX and HP10B3BJEW, 2011 for the availability of high performance computing resources and support. We thank Cormac Reynolds  for his precious technical support and for allowing us to run on CUPPA, the Beowulf cluster of the Curtin Institute of Radio Astronomy (CIRA) at Curtin University. MM and LZ acknowledge financial support from the Italian Ministry of Education, University and Research (MIUR) through grant FIRB 2012 RBFR12PM1F (`New perspectives on the violent Universe: unveiling the physics of compact objects with joint observations of gravitational waves and electromagnetic radiation'), and from INAF through grant PRIN-2011-1 (`Challenging Ultraluminous X-ray sources: chasing their black holes and formation pathways'). 


\end{document}